\documentstyle[twocolumn,aps,psfig,epsfig]{revtex}
\begin{document} 
\draft
\title{Spin fluctuations in a metallic antiferromagnet }
\author{Avinash Singh\cite{avinash} }
\address{Department of Physics, 
Indian Institute of Technology Kanpur - 208016, India}
\maketitle
\begin{abstract} 
The magnon energy and amplitude renormalization due to 
intraband particle-hole excitations are studied in a metallic antiferromagnet.
The change in sign of the intraband contribution with $\omega$ 
results in significant differences between static and dynamical behaviours.
For electron doping, 
while the coherent magnon peak looses spectral weight 
and is shifted to higher energy with doping, 
the low-energy incoherent part becomes increasingly prominent as it 
narrows in width and shifts to lower energy.
Implications for spin dynamics in the electron doped cuprate
$\rm Nd_{2-x}Ce_x Cu O_4$ are discussed.
Due to an exact cancellation of two logarithmically divergent spin-fluctuation processes,
the AF order parameter is unaffected to leading order. 
For hole doping,
short wavelength transverse perturbations ($q > q^* \sim \sqrt{x}$)
are found to be stable, 
implying short-range AF order with a spin correlation length $\xi/a \sim 1\sqrt{x}$. 
\end{abstract}
\section{Introduction}

Antiferromagnetism in the doped cuprates strongly depends on the type of doping.
While the electron doped cuprate $\rm Nd_{2-x}Ce_x Cu O_4$
retains AF order up to a doping concentration of about 
15\%,\cite{electron1,electron2,electron3}
only 2\% hole concentration destroys AF order in $\rm La_{2-x}Sr_x Cu O_4$.

The simplest microscopic description of this strong doping asymmetry
emerges within the $t-t'$ Hubbard model, with nearest-neighbour (NN) and 
next-nearest-neighbour (NNN) hopping terms $t$ and $t'$, respectively. 
For negative $t'$ (as usually assumed for cuprates) and electron doping, 
the AF state is stable for a range of doping concentration,
whereas the AF state becomes unstable 
with respect to transverse perturbations in the AF order 
for any finite hole doping.\cite{chubukov,doped} 
Recently the magnetic phase diagram of the doped $t-t'$ Hubbard model has 
been obtained in the $t'-U$ space, showing the various regions of stability 
and instability with respect to both longitudinal and transverse perturbations 
in the AF order.\cite{doped} 

The appropriate Hubbard model parameters  for $\rm La_{2}CuO_4$,
determined recently by fitting the spin-wave dispersion obtained from high resolution 
inelastic neutron scattering studies,\cite{spinwave,num_rpa,spectrum} 
indicate that $U/t \sim 8$.
From the additional fact that the critical electron doping concentration 
in $\rm Nd_{2-x}Ce_x Cu O_4$ is around $20 \%$, 
the $t'-U$ phase diagram then yields $t'/t \sim 0.25$.\cite{doped}
This value of $t'$ falls in the range 0.15 to 0.5 estimated from 
band structure studies, photoemission data
and neutron-scattering measurements of high-T$_{\rm c}$
and related materials.\cite{nnn1,nnn2,nnn3,nnn4}

A stable AF state of the doped $t-t'$ Hubbard model
provides a microscopic realization of a {\em metallic antiferromagnet},
in which the Fermi energy lies within a quasiparticle band.  
In this paper we will quantitatively study the interplay of 
spin fluctuations and particle-hole excitations in the metallic AF,
focussing on the role of the intraband particle-hole excitations on the 
spin excitation spectrum, magnon renormalization,
and spin-fluctuation correction to the AF order parameter.
These results should be particularly relevant to the 
electron doped cuprate $\rm Nd_{2-x}Ce_x Cu O_4$, 
in which metallic conductivity resulting from electron doping
suggests that the electrons are mobile,\cite{mobile}
and doping with Ce clearly reduces the spin-stiffness constant
from the value in the undoped system.\cite{electron3}
Metallic antiferromagnetism has also been reported in 
$\rm \kappa - (BEDT-TTF)_2 X$,\cite{metaf1}
$\rm V_{2-x} O_3$,\cite{metaf2}  and $\rm NiS_{2-x} Se_x$.\cite{metaf3}

The metallic antiferromagnet is characterized by magnon decay into 
intraband particle-hole excitations, 
and the resulting magnon damping ($\Gamma$) 
was studied recently in the long wavelength limit $(q << 1)$.\cite{doped}
In terms of the Fermi circle radius $a = \sqrt{2\pi x}$
of the doping pockets formed around $(\pm\pi,0)$ and $(0,\pm\pi)$,
where $x$ is the doping concentration,
a new doping-dependent energy scale $4 t' a $ was identified, 
the relative magnitude of which,
in comparison with the magnon energy scale $\sqrt{2}J$,
should essentially determine the spin excitation spectrum.
The imaginary part of $\chi^0({\bf q},\omega)$ was found to vanish 
for $\omega > 4 t' a q$,  
and increase linearly with $\omega$ for $\omega << 4 t' a q$.
This implies that the magnon spectrum should exhibit a sharp peak 
at $\omega \approx \sqrt{2}Jq$ for $\sqrt{2}J > 4t'a$ (low doping limit), 
and magnon broadening should appear with increasing doping
when $\sqrt{2}J < 4t'a$. 
 
In addition to the spin excitation spectrum for all $q$ (section III), 
we will also study the quantum correction to sublattice magnetization
(section IV), to determine whether long-range AF order survives in the 
metallic antiferromagnet when quantum spin fluctuations are included. 
At the one-loop level, the spin-flip process is accompanied by
virtual magnon emission and absorption, 
and therefore we will also examine whether the magnon
propagator (whose amplitude goes like $1/q$ in the AF insulator) 
is significantly renormalized due to the intraband particle-hole
excitations (section II). 
The consequences of magnon renormalization 
on finite temperature spin dynamics and N\'{e}el temperature 
in electron doped cuprates are briefly discussed in section V.

We consider the $t-t'$ Hubbard model on a square lattice,
with NN and NNN hopping terms $t$ and $t'$
connecting sites $i$ to $i+\delta$ and $i+\kappa$, respectively:
\begin{equation}
H = 
-t \sum_{i,\delta,\sigma} ^{\rm NN}
a_{i, \sigma}^{\dagger} a_{i+\delta, \sigma}
-t' \sum_{i,\kappa,\sigma} ^{\rm NNN} 
a_{i, \sigma}^{\dagger} a_{i+\kappa, \sigma}
+  U\sum_{i} n_{i \uparrow} n_{i \downarrow} \; .
\end{equation}
In the following we set $t=1$.
Since a particle-hole transformation maps the $t'$ model with hole (electron)
doping on the $-t'$ model with electron (hole) doping,
the positive (negative) $t'$ model and hole (electron) doping is appropriate to study
for the electron-doped compound $\rm Nd_{2-x}Ce_x Cu O_4$,
for which a negative $t'$ is usually assumed.

\section{Magnon renormalization}
In this section we consider positive $t'$ and hole doping, 
so that the results are 
appropriate for the electron-doped cuprates. 
We also consider the strong coupling limit for analytical simplicity.
To study transverse spin fluctuations in the doped antiferromagnetic state
we evaluate the time-ordered magnon propagator 
\begin{equation}
\chi^{-+}({\bf q},\omega)= \int dt \sum_{{\bf r}_{ij}}
e^{i(\omega t - {\bf q}.{\bf r}_{ij})}
\langle \Psi_{\rm G} | T [ S_i ^- (t) S_j ^+ (0)]|\Psi_{\rm G}\rangle
\end{equation}
involving the spin-lowering and spin-raising 
operators $S_i ^-$ and $S_j ^+$ at lattice sites $i$ and $j$.
In the random phase approximation (RPA), we have
\begin{equation}
\chi^{-+}({\bf q},\omega)= \frac{\chi^0 ({\bf q},\omega)}
{1-U \chi^0 ({\bf q},\omega)}, 
\end{equation}
where $\chi^0 ({\bf q},\omega)$ is the zeroth-order, 
antiparallel-spin particle-hole propagator, 
evaluated in the broken-symmetry, Hartree-Fock (HF) state of the 
metallic antiferromagnet, 
and involves both interband and intraband excitations.
We examine the ${\bf q},\omega$ dependence of $\chi^0 ({\bf q},\omega)$, 
which essentially determines the magnon energy and amplitude renormalization.
We focus on the real part of $[\chi^0 ({\bf q},\omega)]$ in the underdoped limit,
as its imaginary part was shown to vanish for 
$\omega > 4t' a q$ in the long wavelength limit.\cite{doped}

The HF-level description of the metallic AF state has been discussed earlier.\cite{doped} 
The NNN hopping term modifies only the AF-state quasiparticle energies 
$E_{\bf k}$, but not the quasiparticle amplitudes.
We have $E_{\bf k}^{\pm} = \epsilon' _{\bf k} \pm \sqrt{\Delta^2 + \epsilon_{\bf k} ^2}$,
for the upper and lower Hubbard bands, where 
$\epsilon_{\bf k}=-2t(\cos k_x + \cos k_y)$ and 
$\epsilon' _{\bf k} =-4t' \cos k_x \cos k_y$ are the free-fermion energies 
corresponding to the NN and NNN hopping terms. 
Here $2\Delta = mU$ in terms of the sublattice magnetization $m$.

In the two-sublattice basis (labelled by A and B),
the intraband contributions to the diagonal and off-diagonal
matrix elements of $[\chi^0 ({\bf q},\omega)]$ are obtained as:
\begin{eqnarray}
[\chi^0 ({\bf q},\omega)]_{\rm AA} ^{\rm intra}  
&=& 
\frac{1}{4\Delta^2} \sum_{\bf k} ' \left [
\frac
{
\Delta_{\bf q}(\epsilon_{\bf k-q} ^2 + \epsilon_{\bf k} ^2 )  
- \omega (\epsilon_{\bf k-q} ^2 - \epsilon_{\bf k} ^2 ) 
}
{ \Delta_{\bf q} ^2 - \omega^2 }
\right ] \nonumber \\
{[}\chi^{0} ({\bf q},\omega){]}_{\rm BB} ^{\rm intra}
&=&
\frac{1}{4\Delta^2} \sum_{\bf k} ' \left [
\frac
{
\Delta_{\bf q}(\epsilon_{\bf k-q} ^2 + \epsilon_{\bf k} ^2 )  
+ \omega (\epsilon_{\bf k-q} ^2 - \epsilon_{\bf k} ^2 ) 
}
{ \Delta_{\bf q} ^2 - \omega^2 }
\right ] \nonumber \\
{[}\chi^0 ({\bf q},\omega)]_{\rm AB} ^{\rm intra}  
&=&
\frac{1}{4\Delta^2} \sum_{\bf k} ' \left [
\frac
{
\Delta_{\bf q} (2\epsilon_{\bf k} \epsilon_{\bf k-q} )
}
{ \Delta_{\bf q} ^2 - \omega^2 }
\right ] 
=[\chi^0 ({\bf q},\omega)]_{\rm BA} ^{\rm intra}  
\nonumber \\
\end{eqnarray}
where $\sum_{\bf k} '$ indicates that states ${\bf k}$ are below the Fermi
energy $E_{\rm F}$, while states ${\bf k-q}$ are above $E_{\rm F}$,
and $\Delta_{\bf q} \equiv E_{\bf k-q}^\ominus - E_{\bf k}^\ominus$ is 
the intraband particle-hole energy difference in the lower Hubbard band.

The above three intraband contributions can be written,
in units of $(t^2/\Delta^3)$, 
as $(a_{\bf q}-b_{\bf q}\omega/2J)$, 
$(a_{\bf q}+b_{\bf q}\omega/2J)$, and $c_{\bf q}$, respectively.
The dimensionless coefficients $a_{\bf q},b_{\bf q},c_{\bf q}$ 
are all functions of $\omega^2$ and, for $\omega \ne 0$, 
vanish quadratically with $q$ as $q \rightarrow 0$ due to phase space 
restriction.\cite{doped}
However, for $\omega \propto q$ (near the magnon-mode energy),
they all have well defined limits as $q \rightarrow 0$.

Including the interband contributions as well,
which were studied earlier for finite doping up to order $x^2$,\cite{doped}
$[\chi^0 ({\bf q},\omega)]$ can be written as 
\begin{equation}
\chi^0 ({\bf q},\omega)= \frac{1}{U} - \frac{t^2}{\Delta ^3}
\left [ \begin{array}{lr} 
1- a' _{\bf q} - a_{\bf q} + \tilde{\omega} & \gamma_{\bf q} - c_{\bf q} \nonumber \\
\gamma_{\bf q} - c_{\bf q} & 1- a' _{\bf q} - a_{\bf q} - \tilde{\omega}
\end{array}
\right ]\; .
\end{equation}
The term $a' _{\bf q}$ represents modifications in the interband contribution 
due to NNN exchange energy $J'=4t^{'2}/U$ and finite doping, 
and to first order in $x$, is given by
\begin{eqnarray}
a' _{\bf q} &=&
\frac{J'}{J}\left (1-\frac{\Delta x}{t'}\right ) 
(1-\gamma' _{\bf q})  \nonumber \\
&+&  x (\cos q_x -\cos q_y)^2  
-  \frac{2J'}{J} \, x (1-\gamma' _{\bf q} )^2  \; ,
\end{eqnarray}
where $\gamma_{\bf q}=(\cos q_x + \cos q_y)/2$ and 
$\gamma' _{\bf q}=\cos q_x \cos q_y)$.
The scaled frequency $\tilde{\omega}=m^2(1+b_{\bf q})(\omega/2J) \equiv 
f_{\bf q} \, (\omega/2J)$, where $m=1-x$, reflects the frequency-scale renormalization
due to doping and intraband excitations.

Substituting $\chi^0 ({\bf q},\omega)$ from Eq. (5) in Eq. (3),
the RPA-level magnon propagator is obtained as 
\begin{eqnarray}
& & \chi^{-+}({\bf q},\omega) = - \frac{1}{2}\frac{m}{1+b_{\bf q}} f_{\bf q}^{-1}
\nonumber \\
&\times & \left [ \begin{array}{lr} 
(1- a' _{\bf q} - a_{\bf q}) - \tilde{\omega} 
& - (\gamma_{\bf q} - c_{\bf q}) \nonumber \\
\nonumber \\
- (\gamma_{\bf q} - c_{\bf q}) & (1- a' _{\bf q} - a_{\bf q}) 
+ \tilde{\omega}
\end{array}
\right ]\; \nonumber \\
&\times & 
\frac{2J}{\Omega_{\bf q}(\omega)}
\left (
\frac{1}{\omega - \Omega_{\bf q}(\omega) } 
- \frac{1}{\omega + \Omega_{\bf q}(\omega) } \right ) \; ,
\end{eqnarray}
where
\begin{equation}
\Omega_{\bf q}(\omega) = 
2J f_{\bf q}^{-1}
[( 1 - a'_{\bf q} - a_{\bf q} )^2 - (\gamma_{\bf q} -c_{\bf q})^2 ]^{1/2}
\end{equation}
is frequency dependent due to the intraband contribution, 
resulting in the magnon amplitude renormalization. 
The magnon-mode energies for momentum $q$ are given by
the poles at $\omega_{\bf q} = \pm \Omega_{\bf q}(\omega_{\bf q})$. 

For $J << 8t'$, as in the strong coupling limit, 
the magnon renormalization due to $b_{\bf q}$ is negligible,
and in the following we take the frequency-scale renormalization factor 
$f_{\bf q} = m^2 (1+b_{\bf q}) \approx 1 $, for simplicity. 

In the long-wavelength and low-doping limits ($q<<1, x<<1$)
we have $\gamma_{\bf q} \approx 1-q^2/4$, 
$\gamma '_{\bf q} \approx 1-q^2/2$, and $a_{\bf q},b_{\bf q},c_{\bf q} << 1$.
Substituting for $a'_{\bf q}$ from Eq. (6) in Eq. (8), 
the expression for $\Omega_{\bf q}(\omega)$ simplifies to 
\begin{equation}
\Omega_{\bf q}(\omega) = \sqrt{2}J 
[\alpha_{\rm inter} - \alpha_{\rm intra}(\omega)]^{1/2} q
\end{equation}
where the dimensionless quantities 
\begin{eqnarray}
\alpha_{\rm inter} &=& 1-(2J'/J) (1-\Delta x/t')  \\
{\rm and \;\;} \alpha_{\rm intra}(\omega) &=& 4(a_{\bf q}-c_{\bf q})/q^2 
\end{eqnarray}
were introduced earlier as coefficients of
the $q^2$ term in the interband and intraband contributions to the 
transverse response eigenvalue in the context of the 
stability analysis of the doped AF state of 
the $t-t'$ Hubbard model.\cite{doped}
From Eq. (4) which defines $a_{\bf q}$, $b_{\bf q}$, and $c_{\bf q}$, 
and substituting for $\epsilon_{\bf k-q} - \epsilon_{\bf k} \approx 2tqa \cos\theta$,  
and the intraband particle-hole energy difference 
$E_{\bf k-q}^\ominus - E_{\bf k}^\ominus \equiv \Delta_{\bf q}
\approx 4t'q a \cos \theta$ in the limits $t' > Jx$ and $q << 1$,
we obtain 
\begin{equation}
\alpha_{\rm intra}(\omega) 
= \frac{\Delta x }{t'} \int_{-\pi/2} ^{\pi/2}
\frac{d\theta}{\pi}
\frac{\cos^4 \theta}{\cos^2 \theta - (\omega/4t'qa)^2 } \; .
\end{equation}
As $\alpha_{\rm intra}(\omega)$ is a function of $\omega/q$,
it follows from Eq. (9) that the magnon-mode energy
$\omega_{\bf q}$ is proportional to $q$, as in the AF insulator.

The change in sign of the intraband coefficient 
$\alpha_{\rm intra}(\omega)$ at $\omega/4t' q a =1$ (see Fig. 1) has interesting
consequences on the doping dependence of the magnon-mode energy.
In the static limit, $\alpha_{\rm intra}(0) = \Delta x /2t' $ is positive, 
and the intraband contribution softens the magnon mode,
eventually leading to the instability of the AF state when 
$\alpha_{\rm intra} = \alpha_{\rm inter}$.\cite{doped}
However, for $\omega > 4t' qa$, the intraband coefficient $\alpha_{\rm intra}$ 
is negative, implying that the intraband excitation actually stiffens the magnon-mode. 

Therefore, for $\sqrt{2}J > 4t'a = 4t'\sqrt{2\pi x}$ (underdoped limit), 
the coherent magnon peak at $\omega_{\bf q} \approx \sqrt{2}Jq$ is shifted to higher energy,
this shift being proportional to $x^2$ in the small doping limit.
With increasing doping, when $\sqrt{2}J <  4t'a $,   
the magnon peak is broadened due to damping and gets shifted to lower energy
due to the change in sign of $\alpha_{\rm intra}(\omega)$. 
As the magnon damping term goes as $\omega$ for $\omega << 4t'qa$,\cite{doped}
the magnon peak significantly narrows down as it shifts to lower energies with doping.

Doping stiffens the magnon mode through the interband contribution as well,
by suppressing the softening due to the $J'$-induced frustration.
This is clearly seen from the expression for the interband coefficient
$ \alpha_{\rm inter}$ in Eq. (10), which increases with $x$. 

\begin{figure}
\vspace*{-70mm}
\hspace*{-38mm}
\psfig{file=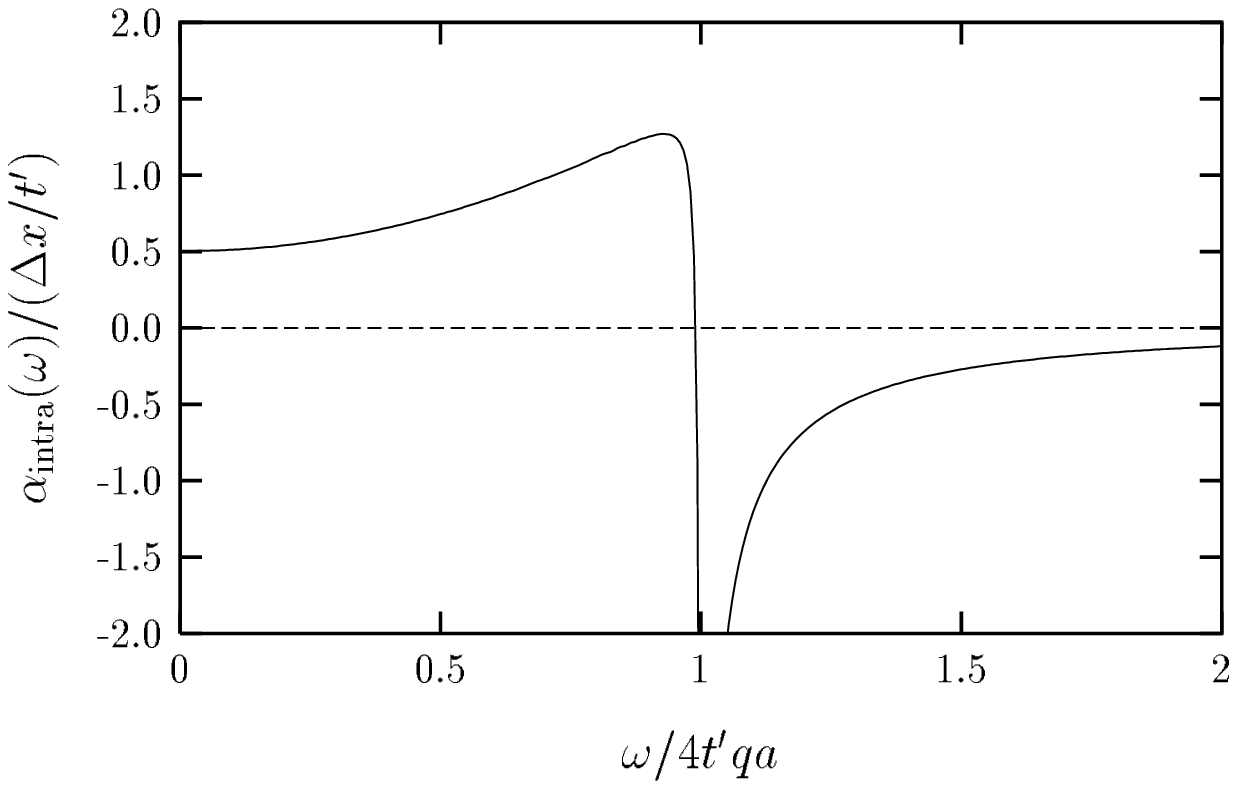,width=135mm,angle=0}
\vspace{-70mm}
\caption{The intraband coefficient $\alpha_{\rm intra}(\omega)$ changes sign
at $\omega = 4t'qa$, implying that the magnon-mode softening for 
$\omega_{\bf q} < 4t'qa$ changes into magnon-mode stiffening for 
$\omega_{\bf q} > 4t'qa$.}
\end{figure}

Turning now to the magnon amplitude renormalization,
by expanding the frequency-dependent energy 
$\Omega_{\bf q}(\omega)$ near the mode energy $\omega_{\bf q}$,
we obtain 
\begin{equation}
\omega - \Omega_{\bf q}(\omega) 
\approx \left ( 1- \left.\frac{\partial\Omega_{\bf q}}{\partial\omega}\right |
_{\omega_{\bf q}} \right ) (\omega - \omega_{\bf q}) \; ,
\end{equation}
which yields the magnon renormalization factor 
\begin{equation}
Z = \left ( 1- \left. \frac{\partial\Omega_{\bf q}}{\partial\omega} \right |
_{\omega_{\bf q}} \right )^{-1}.
\end{equation}
From Eqs. (9) and (12), we obtain
\begin{equation}
\left. \frac{\partial\Omega_{\bf q}}{\partial\omega} \right |_{\omega_{\bf q}} =
- \left (\frac{\sqrt{2}J}{4t'a}\right)^2 \frac{\Delta x}{t'} 
\int_{-\pi/2} ^{\pi/2} \frac{d\theta}{\pi}
\frac{\cos^4 \theta}{[ \cos^2 \theta - (\omega_{\bf q}/4t'qa)^2  ]^2 }
\end{equation}
which is negative, and vanishes with the doping concentration as $x^2$,
through the $x$ dependence of the integral.
These consequences of doping,
namely the stiffening of the magnon mode and 
reduction in the coherent spectral weight  for low doping, 
changing into magnon softening and broadening with increasing doping,
are confirmed in a numerical study of the magnon spectral function, as discussed below.

\section{Spin fluctuation spectrum}
The spectrum of transverse spin fluctuations in the doped AF state,
evaluated numerically from the imaginary part of the magnon propagator 
$\chi^{-+}({\bf q},\omega)$, is discussed in this section.
In terms of the two complex eigenvalues $\lambda_n({\bf q},\omega)$ 
of the $\chi^0 ({\bf q},\omega)$ matrix, we obtain for the 
spin fluctuation spectral function $A_{\bf q}(\omega)$

\begin{figure}
\vspace*{-70mm}
\hspace*{-38mm}
\psfig{file=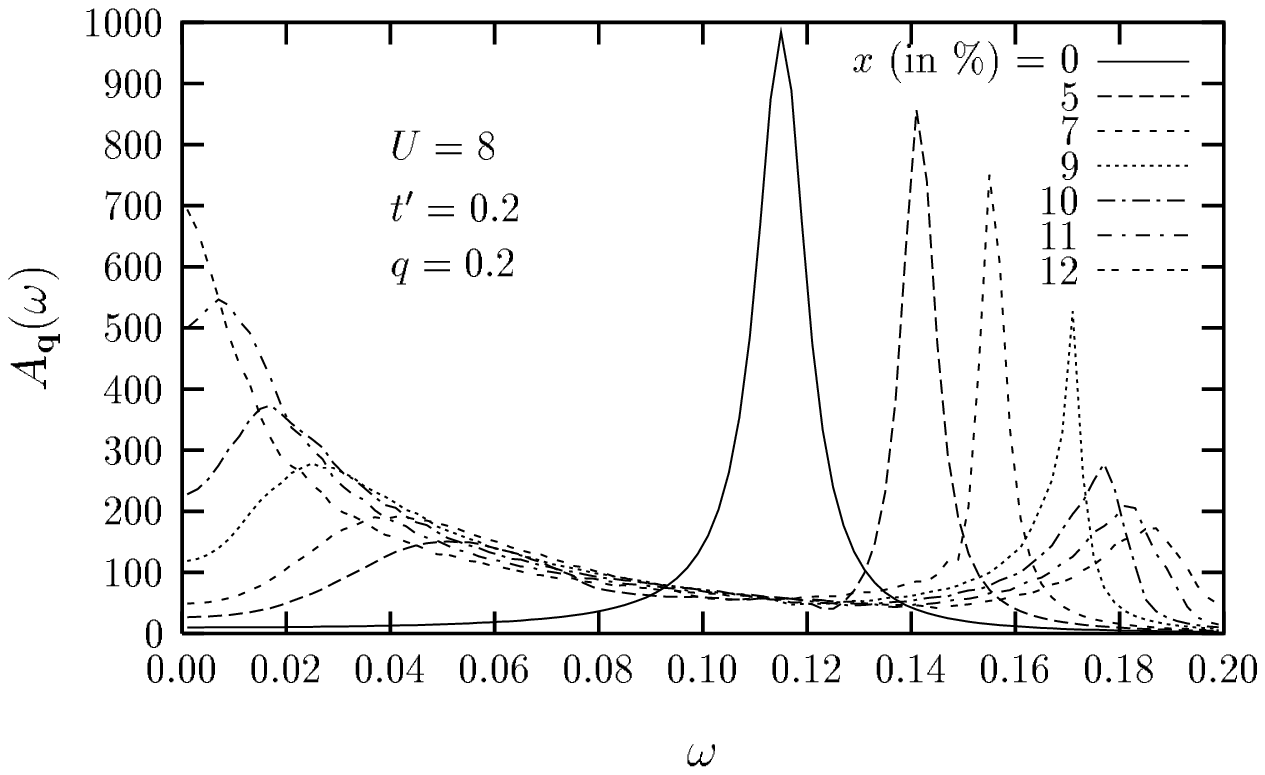,width=135mm,angle=0}
\vspace{-70mm}
\caption{
The magnon spectral function 
$A_{\bf q}(\omega) = {\rm Tr} \; {\rm Im}\; \chi^{-+}({\bf q},\omega) $
for a long wavelength mode ${\bf q}=(0.2,0)$, showing the shift of the coherent 
magnon peak to higher energy with doping,
accompanied by a reduction of the magnon amplitude.
With increasing doping the broad incoherent spectrum acquires strength
and narrows into a peak which shifts down in energy.}
\end{figure}

\begin{figure}
\vspace*{-70mm}
\hspace*{-38mm}
\psfig{file=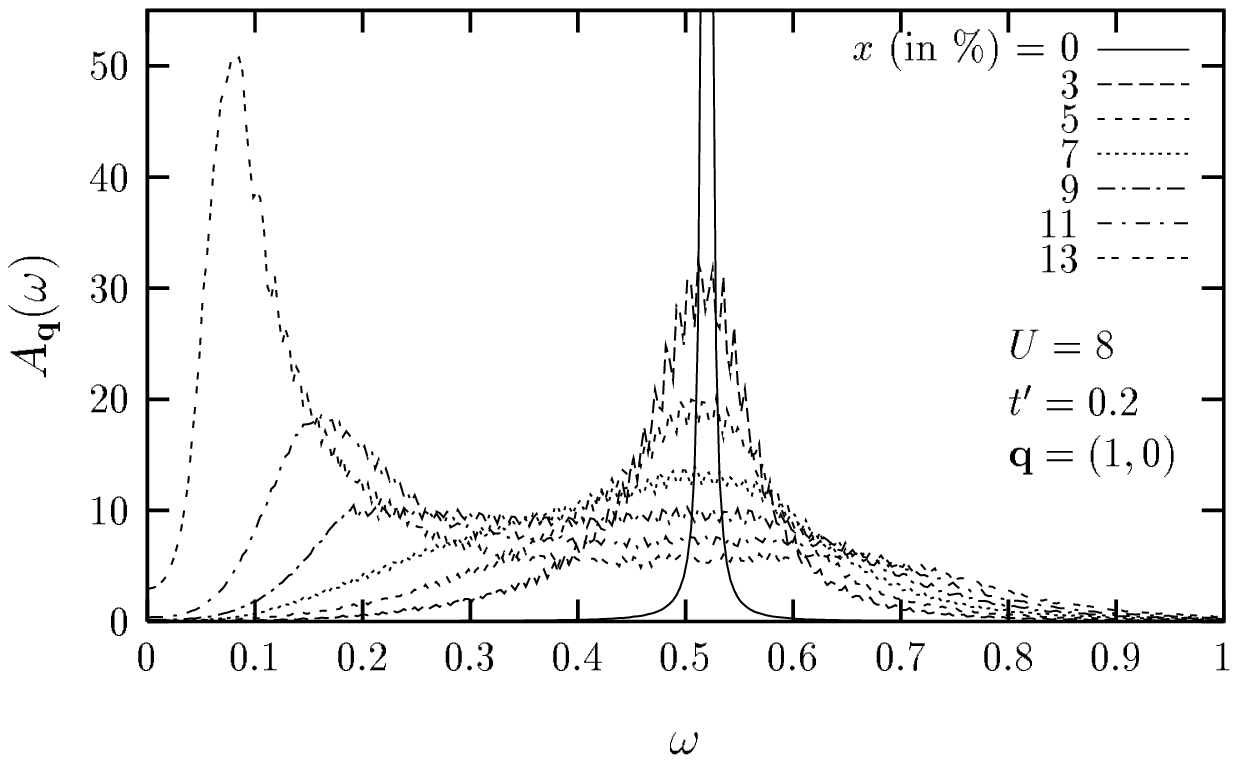,width=135mm,angle=0}
\vspace*{-70mm}
\caption{
For the short-wavelength mode ${\bf q}=(1,0)$,
the magnon peak broadens almost immediately with doping,
with no change in the peak energy.
With increasing doping, the magnon peak becomes strongly
asymmetric, rapidly narrowing into a strong peak at low energy.}
\end{figure}

\begin{equation}
A_{\bf q}(\omega) = {\rm Tr} \; {\rm Im}\; \chi^{-+}({\bf q},\omega) 
= \sum_{n=1,2} {\rm Im} \frac{\lambda_n({\bf q},\omega)}
{1-U\lambda_n({\bf q},\omega)} \; .
\end{equation}
In the numerical evaluation of the $\chi^0 ({\bf q},\omega)$ matrix, 
we have taken a grid size $dk_x=dk_y =0.01$
and an imaginary term $\eta=0.001$ throughout. 
The spectral function $A_{\bf q}(\omega)$ has been recently studied 
for the AF insulator in the full $U$ range from weak coupling to strong coupling,
including contributions from both single-particle 
and collective (magnon) excitations.\cite{spectral}
In the following, 
we discuss the results in the context of cuprates,
separately considering electron doping (positive $t'$ and hole doping) 
and hole doping (negative $t'$ and hole doping).

\begin{figure}
\vspace*{-70mm}
\hspace*{-38mm}
\psfig{file=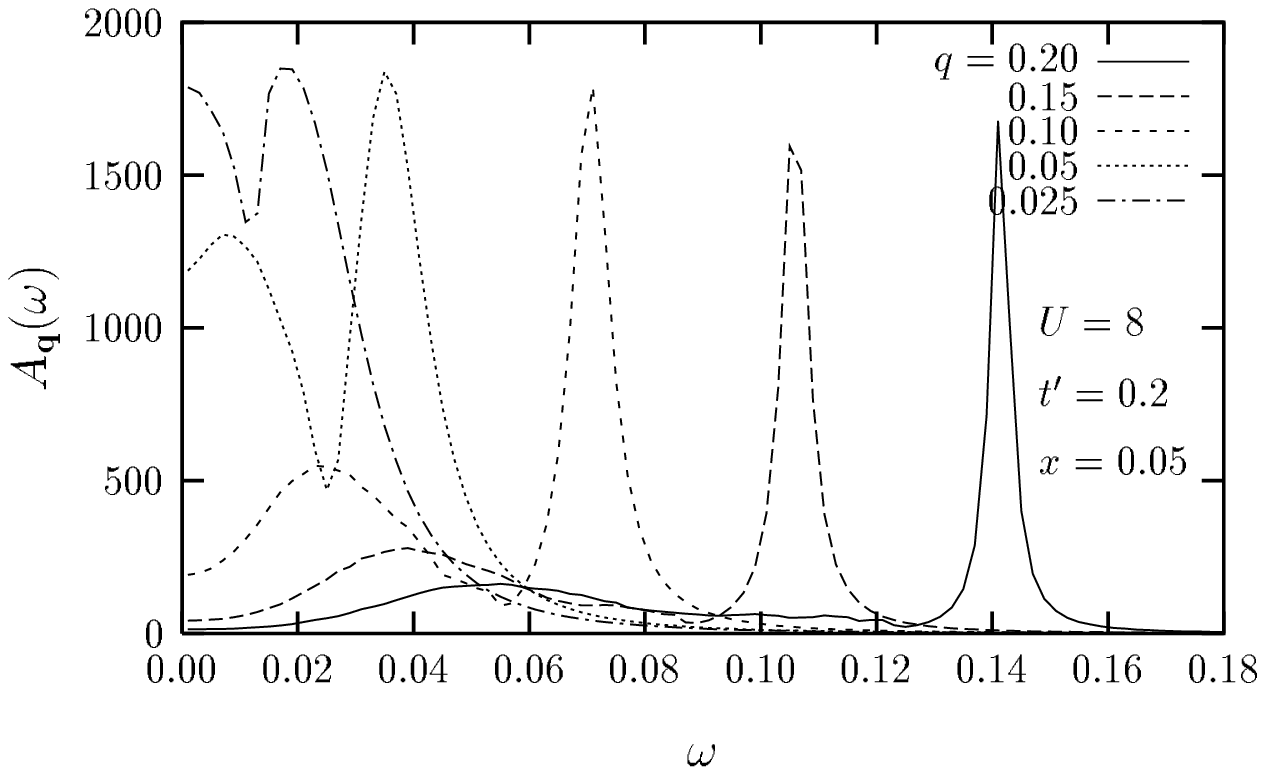,width=135mm,angle=0}
\vspace{-70mm}
\caption{
For a fixed doping concentration,
the magnon spectral function for different $q$ shows a linear dependence 
of the coherent magnon peak energy $\omega_{\bf q}$ on $q$.
With decreasing $q$ the incoherent part is compressed
into a narrow peak, which competes in strength with the 
coherent peak.}
\end{figure}

\subsection{Electron doping}
The results for $A_{\bf q}(\omega)$ are shown in Figs. 2-5, 
which confirm the characteristic features
obtained in the magnon renormalization study of section II in the
small $q$ limit. 
With doping, the coherent magnon peak 
shifts to higher energy with decreasing amplitude (see Fig. 1).
Simultaneously, the incoherent part of the magnon 
spectral function at low frequency progressively becomes narrower, 
shifts to lower energy and develops oscillator strength.
The softening is due to the intraband contribution and the
narrowing follows from the behaviour of 
${\rm Im} \; \chi^0 ({\bf q},\omega)$, which decreases with $\omega$. 

On the other hand, for short wavelength modes, 
the magnon peak is seen to broaden almost immediately with 
doping (see Fig. 3), with no change in the peak energy.
A study of the imaginary part of $\chi^0({\bf q},\omega)$ shows that 
the energy range over which ${\rm Im} \chi^0({\bf q},\omega)$ is finite 
initially increases as $4t'aq$, 
but then increases more rapidly with $q$, eventually overtaking the magnon energy. 
Therefore for short wavelength modes, 
the magnon energy falls within this range,
resulting in the broadening.

The magnon softening is seen to be moderately $q$ dependent.
While the magnon peak shifts to nearly $\omega \approx 0$ at $x \sim 10 \%$
for  small $q$, a substantial energy gap still 
remains for $q \sim 1$ (see Fig. 3).
This interestingly shows that the instability of the AF state
with respect to transverse fluctuations in the order parameter
strongly depends on the fluctuation wavelength. 
While the AF state becomes unstable with respect to long
wavelength modes at $x = x_c$, the short wavelength modes are
still positive energy modes, so that short-range AF order
should survive.

\begin{figure}
\vspace*{-70mm}
\hspace*{-38mm}
\psfig{file=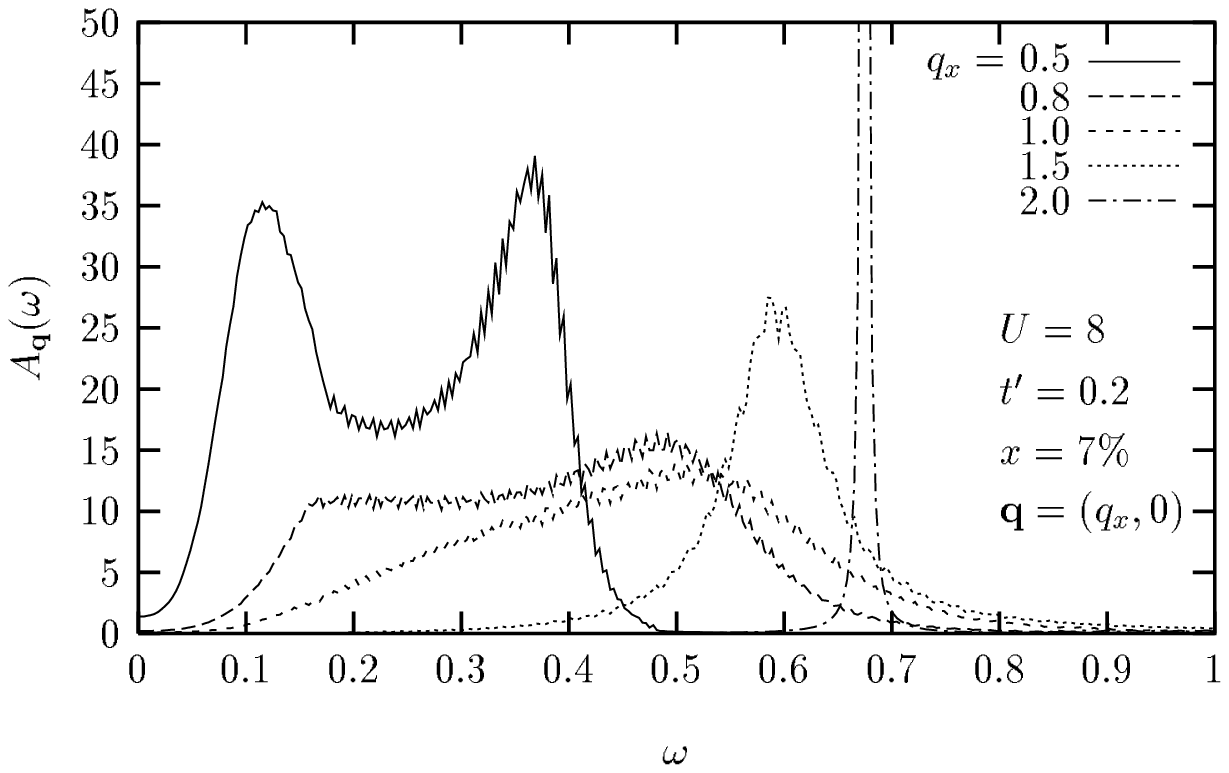,width=135mm,angle=0}
\vspace{-70mm}
\caption{
The incoherent part of the magnon spectral function decreases progressively
with increasing wavevector ${\bf q}=(q_x,0)$,
for a fixed doping concentration of $x=7\%$.
Magnon damping and linewidth sharply decrease as $q_x$ approaches $\sim 2.0$, 
and for $q_x \ge 2$ only the coherent magnon peak remains at low energy,
with an insignificant incoherent part at higher energy (not shown).}
\end{figure}

For a fixed doping concentration, the coherent magnon peak energy is proportional to
the wavevector $q$ in the small $q$ limit (see Fig. 4).
For short wavelength modes, 
the incoherent part decreases dramatically with increasing wavevector (see Fig. 5).
Magnon damping and linewidth sharply decrease as $q$ approaches $\sim 2.0$, 
and for $q \ge 2$ only the coherent magnon peak remains at low energy,
with an insignificant incoherent part at higher energy.

As the doping approaches the critical concentration $x_c$,
above which the AF state is unstable,
the appearance of narrow magnon modes at very low energy
despite their strong renormalization due to the intraband particle-hole
excitations, is a noteworthy feature of the metallic AF state.

\subsection{Hole doping}
The strong intraband contribution renders the 
homogeneous AF state unstable for any amount of hole doping.\cite{chubukov,doped} 
This instability with respect to transverse perturbations 
in the AF order is signalled by the transverse response eigenvalue
$U\lambda_{\bf q}$ exceeding unity for long wavelength (small $q$) modes, 
indicating absence of long-range AF order,
and signalling a tendency towards incommensurate ordering with wavevector different from
${\bf Q}=(\pi,\pi)$. Different types of homogeneous spiral phases,
and their stability with respect to longitudinal and transverse perturbations
have been studied in detail.\cite{chubukov} 

However, short-range AF order appears stable,
as indicated by the full $q$ dependence of $U\lambda_{\bf q}$ (see Fig. 6)
for the two eigenvalues $\lambda_{\bf q}$ of the $\chi^0({\bf q})$ matrix
for ${\bf q}=(q,q)$ in the range $0 < q < \pi $.
The reflection symmetry about 

\begin{figure}
\vspace*{-70mm}
\hspace*{-38mm}
\psfig{file=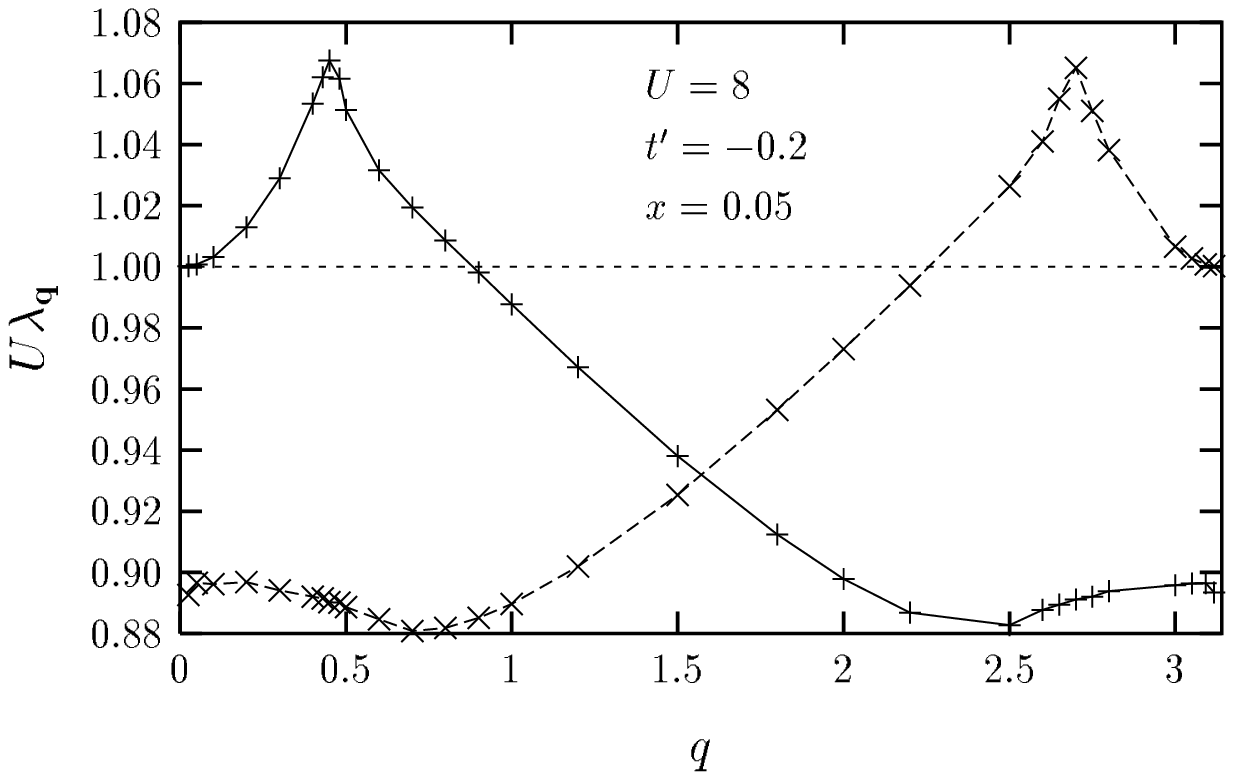,width=135mm,angle=0}
\vspace{-70mm}
\caption{The $q$ dependence of $U\lambda_{\bf q}$ 
in the range $0 \le q \le \pi $ 
for the two eigenvalues $\lambda_{\bf q}$ of the $\chi^0({\bf q})$ matrix
for hole concentration of $5\%$.}
\end{figure}

\begin{figure}
\vspace*{-70mm}
\hspace*{-38mm}
\psfig{file=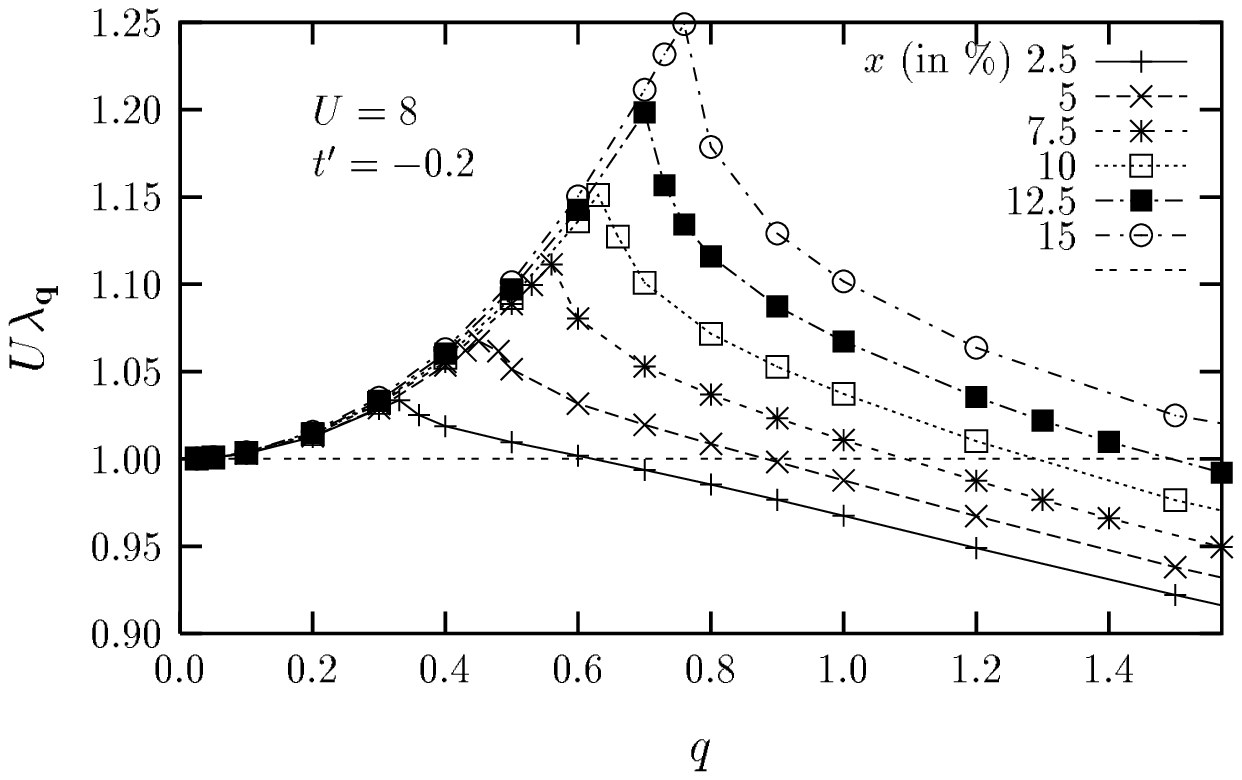,width=135mm,angle=0}
\vspace{-70mm}
\caption{The $q$ dependence of $U\lambda_{\bf q}$ for different hole doping 
concentrations $x$, showing that $U\lambda_{\bf q}$ drops below 1 
above some value $q=q^*$. Note that the small-$q$ parabolic behaviour 
of $U\lambda_{\bf q}$ is independent of doping concentration.}
\end{figure}

\noindent
$q=\pi/2$ is due to the 
location of the hole pockets around $(\pm\pi/2,\pm\pi/2)$ in the Brillouin zone.
Considering the larger eigenvalue for small $q$, 
the parabolic behaviour of $U\lambda_{\bf q}$ is independent of doping
concentration (see Fig. 7), as obtained earlier.\cite{doped}
However, beyond a certain $q$ value $U\lambda_{\bf q}$ starts decreasing
and eventually drops below unity at $q=q^*$.
This signals the stability of the AF state with respect to transverse perturbations 
of wavelength shorter than $\lambda^*=2\pi/q^*$,
implying that short-range AF order can exist
in the hole doped AF upto length scale $\xi \sim \lambda^*$. 
For $x \ge 15\%$, the transverse response eigenvalue $U\lambda_{\bf q} > 1 $ for all $q$, indicating no stability for AF domains of any size.

The $q$ dependence of $U\lambda_{\bf q}$ for different hole doping concentrations
(Fig. 7) shows that the wavevector value $q^*$ where 
$U\lambda_{\bf q}$ drops below unity increases with doping concentration $x$,
indicating diminishing spin correlation length.
The behaviour of the spin correlation length $\xi/a = 2\pi/q^*$ with $x$
is shown in Fig. 8, clearly showing a $1/\sqrt{x}$ dependence,
as seen in neutron scattering experiments.\cite{spincor}

\begin{figure}
\vspace*{-70mm}
\hspace*{-38mm}
\psfig{file=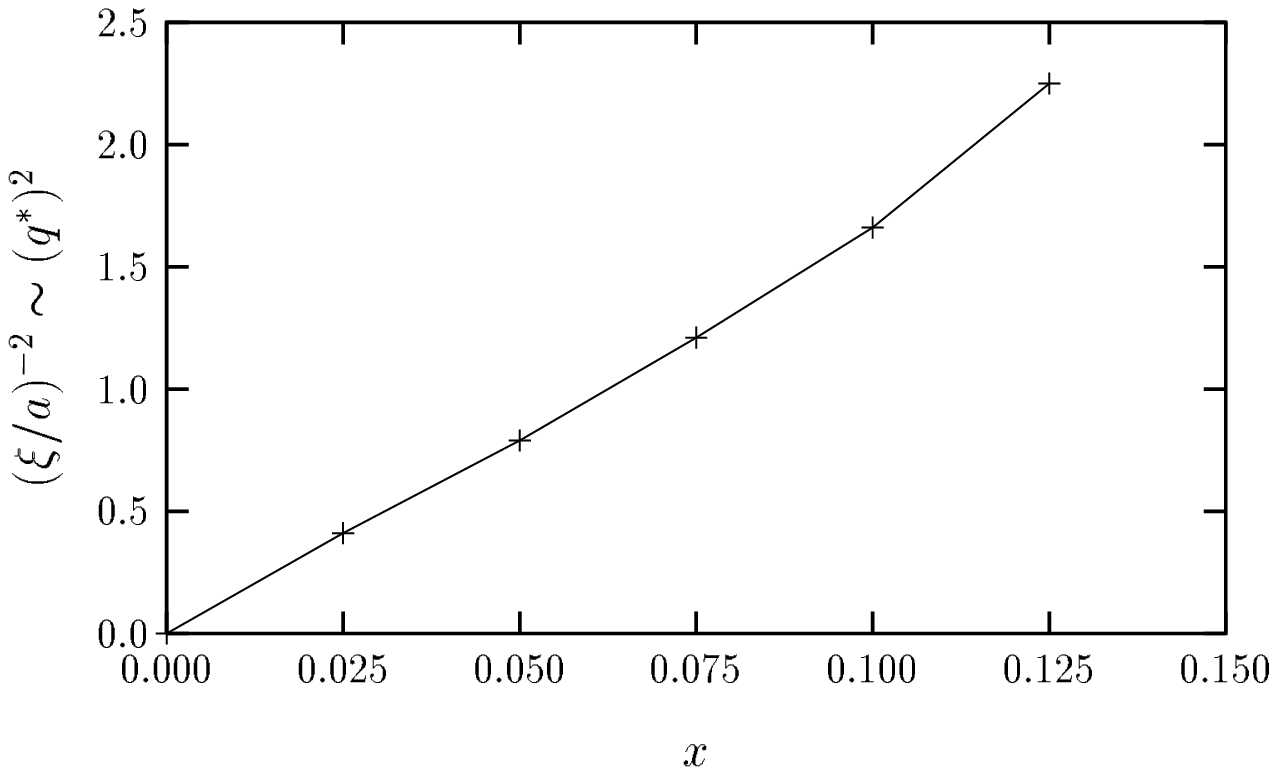,width=135mm,angle=0}
\vspace{-70mm}
\caption{
Plot of $(q^*)^2$ vs. $x$ shows a linear dependence,
implying a $x^{-1/2}$ dependence of the spin correlation length $\xi$ 
on hole doping concentration $x$.}
\end{figure}

\begin{figure}
\vspace*{-70mm}
\hspace*{-38mm}
\psfig{file=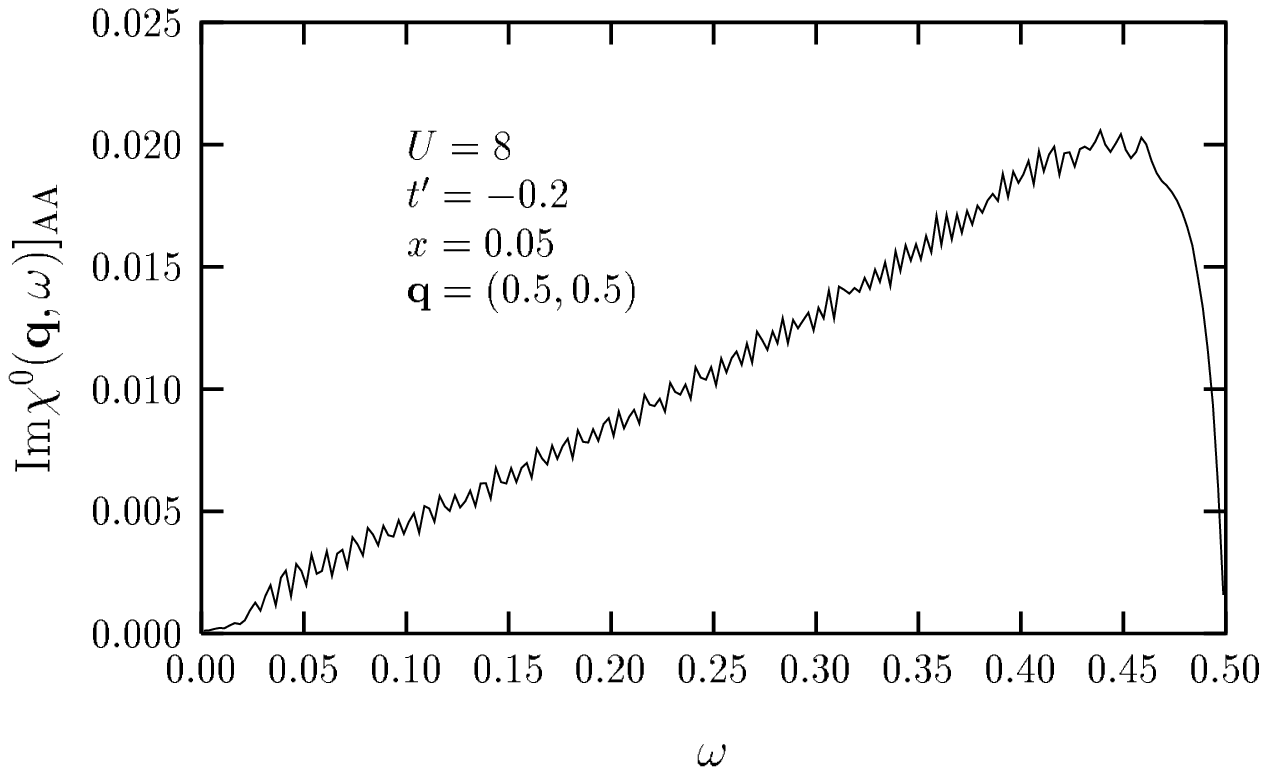,width=135mm,angle=0}
\vspace{-70mm}
\caption{The imaginary part of the particle-hole 
propagator $\chi^0 ({\bf q},\omega)$ shows a linear behaviour 
over a large $\omega$ range.}
\end{figure}

Figure 9 shows that imaginary part ${\rm Im} [ \chi^0 ({\bf q},\omega) ]_{\rm AA}$
on the A sublattice has a remarkably linear behaviour over a large $\omega$ range.
A similar behaviour for $q$ not too small is also observed in the electron-doped case.
This low-frequency  contribution in ${\rm Im} \chi^0 ({\bf q},\omega)$,
arising from the intraband excitations in the antiparallel-spin particle-hole
propagator $\chi^0 ({\bf q},\omega)$, 
is suppressed to order $t^2/U^2$ due to the AF coherence factors.   
However, no such suppression is present in the parallel-spin particle-hole propagators  
$[\pi^0 ({\bf q},\omega)]_{\rm AA}$ or $[\pi^0 ({\bf q},\omega)]_{\rm BB}$, 
where both particle and hole amplitudes (in the lower band) are of order unity.
Such parallel-spin particle-hole propagators 
are involved in the charge density fluctuations.

Although the AF state is unstable at the static level, 
the spectral function shows sharp peaks at finite frequency (see fig. 10).
This is due to the change in sign of the intraband coefficient 
$\alpha_{\rm intra}(\omega)$ with increasing $\omega$, discernible from the change in
curvature of the spectral function. As for the electron-doped case, 
the small-$q$ magnon peaks are sharp because 
${\rm Im} \chi^0 ({\bf q},\omega) $ vanishes at the magnon-mode energy,
whereas at higher $q$ the magnon-mode energy lies within the energy scale 
over which ${\rm Im} \chi^0 ({\bf q},\omega) $ is finite 

\begin{figure}
\vspace*{-70mm}
\hspace*{-38mm}
\psfig{file=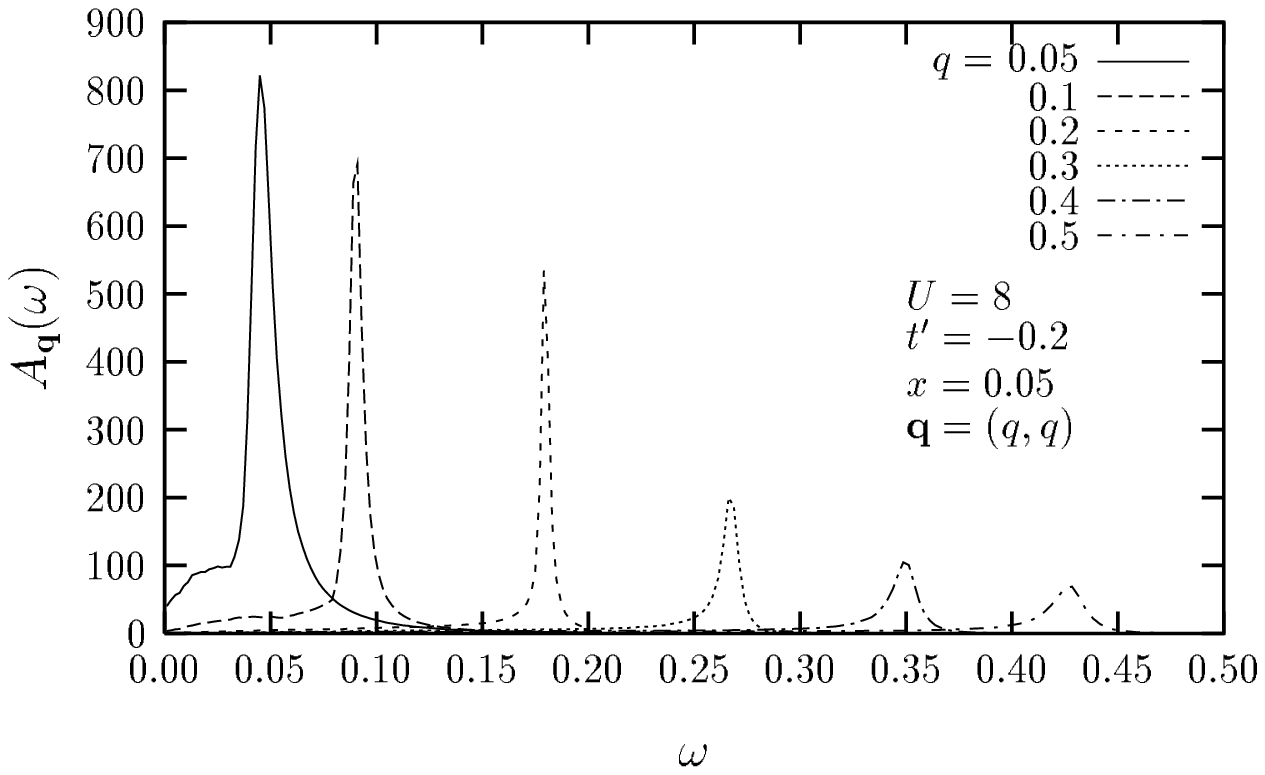,width=135mm,angle=0}
\vspace{-70mm}
\caption{
The spectral function shows sharp magnon peaks for low $q$, 
and broadened peaks at higher $q$ due to magnon damping.}
\end{figure}

\begin{figure}
\vspace*{-70mm}
\hspace*{-38mm}
\psfig{file=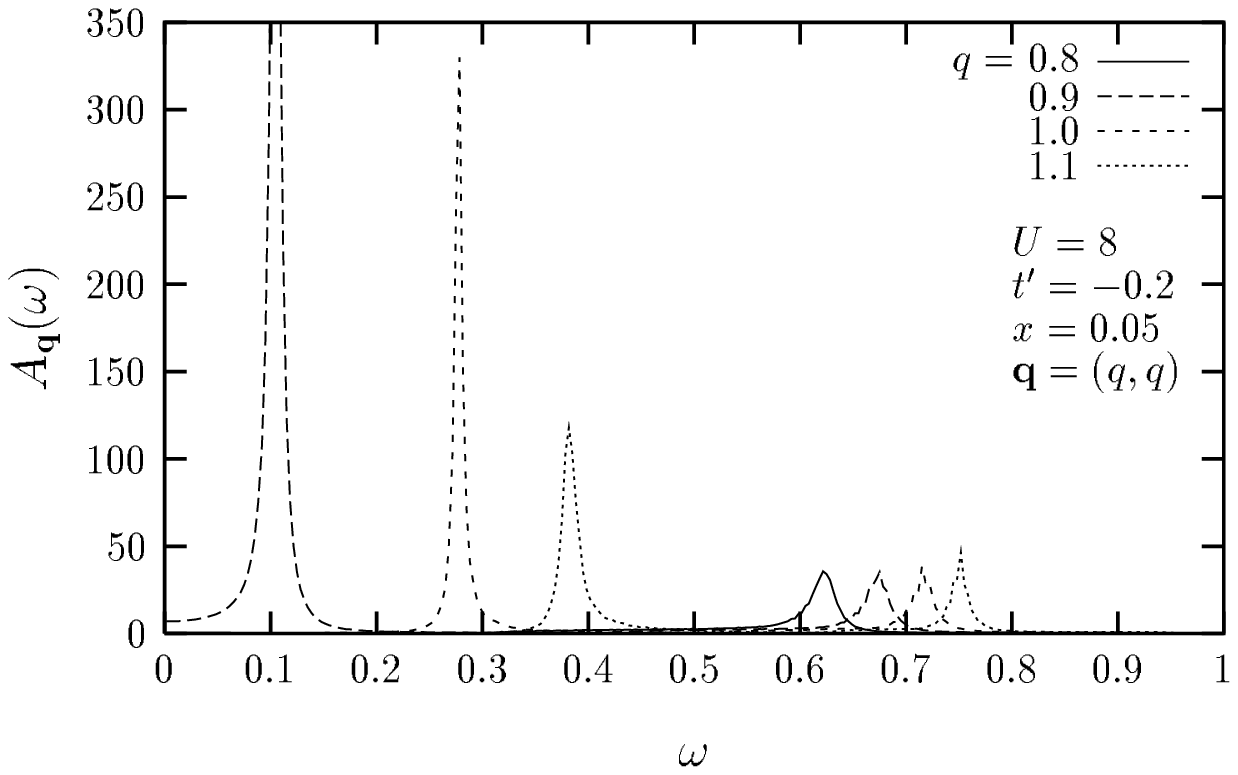,width=135mm,angle=0}
\vspace{-70mm}
\caption{
For fluctuation modes of wavelength smaller than the domain size,
($q > q^* \approx 0.9$ for $x=0.05$), 
the magnon modes are strongly softened,
with a small incoherent part appearing at {\em higher} energy.}
\end{figure}

\noindent
(see Fig. 9), resulting in magnon broadening. 

Figure 11 shows that there is an abrupt change in the spectral function
when $q$ crosses $q^* \approx 0.9$ for $x=0.05$ (see Figure 7).
For fluctuation modes of wavelength smaller than 
the AF domain size ($q > q^*$),
the magnon modes are strongly softened by the intraband contribution,
with a small incoherent part appearing at {\em higher} energy. 
In contrast, in the electron-doped case, the coherent part was shifted to
higher energy and a weak incoherent part appeared at low energy. 
Both magnon damping  and softening have been observed in
the hole-doped cuprate  $\rm La_{2-x} Ba_x Cu O_4$.\cite{aeppli_89}

\section{Finite temperature spin dynamics}
At finite temperature $T<<J$,
the decrease in sublattice magnetization $m(T)$ due to thermal excitation 
of long wavelength magnons
has been studied in a highly anisotropic layered antiferromagnet
with planar exchange 

\begin{figure}
\vspace*{-70mm}
\hspace*{-38mm}
\psfig{file=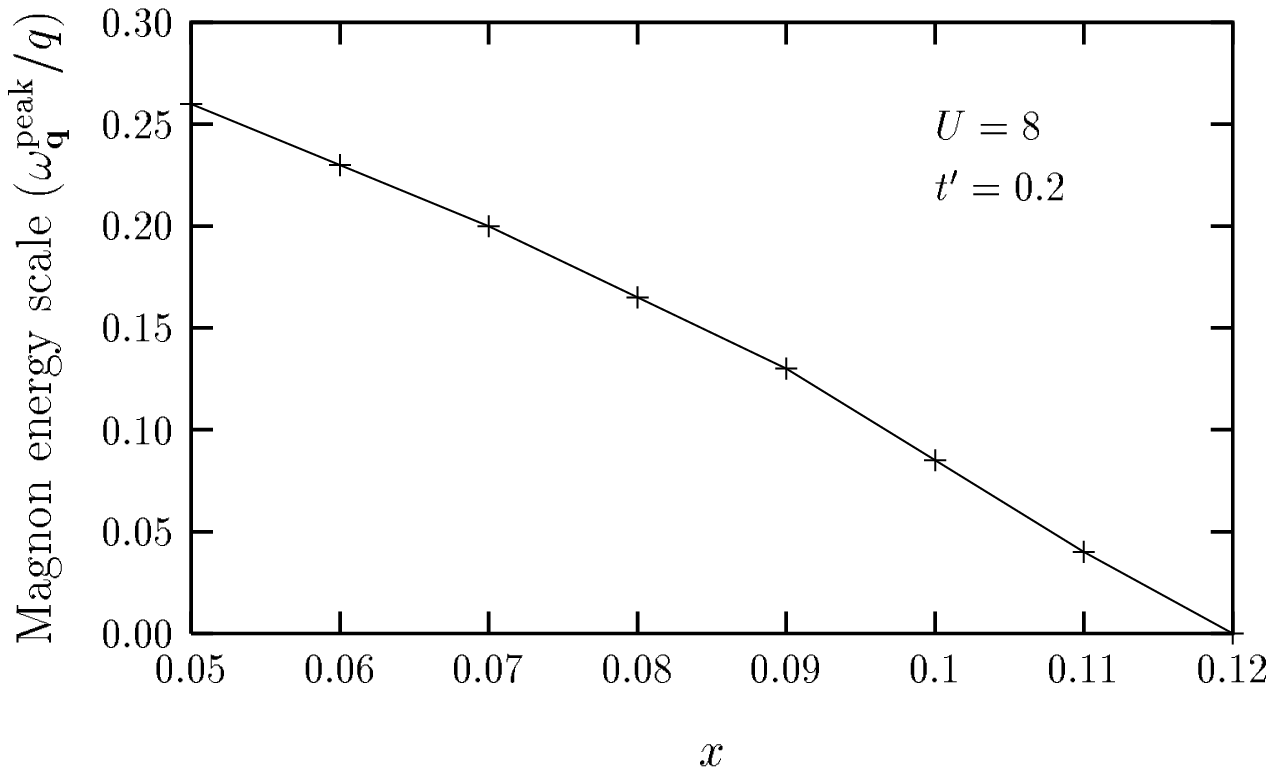,width=135mm,angle=0}
\vspace{-70mm}
\caption{
The long-wavelength magnon energy scale $\omega_{\bf q}^{\rm peak}/q$ 
falls linearly with the doping concentration $x$,
implying a corresponding decrease in the spin-wave stiffness constant
and the N\'{e}el temperature.}
\end{figure}

\noindent
interaction energy $J=4t^2/U$, 
an effective interlayer hopping $t_z$,\cite{prl}
and an effective anisotropy gap $\Delta_{\rm DM}$ 
due to the Dzyaloshinski-Moriya interaction.\cite{anisotropy}
An estimate of the N\'{e}el temperature $T_{\rm N}$ was obtained from 
the predominant $T\ln T$ fall off of $m(T)$ with $T$ as  
$k_{\rm B} T_{\rm N} \sim J/ \ln (1/r)$ for $\Delta_{\rm DM} << 2Jr$, and 
$k_{\rm B} T_{\rm N} \sim J/ \ln (J/\Delta_{\rm DM})$ for $2Jr << \Delta_{\rm DM}$,
where $r=t_z/t$ is the interlayer-to-planar hopping ratio.
In both cases, $T_{\rm N}$ is proportional to the 
long wavelength magnon energy scale  $(\omega_{\bf q}/q)_{q\rightarrow 0} \sim J$.

Figure 2 shows that the relevant magnon energy scale which will 
essentially determine the low temperature spin dynamics in the 
doped AF is the incoherent magnon peak energy, 
which shifts to lower energy and gains spectral weight with increasing doping.
The doping dependence of this magnon energy scale $\omega_{\bf q}^{\rm peak}/q$ 
shows a nearly linear decrease with doping concentration (see Fig. 12).

This nearly linear decrease in the 
incoherent magnon peak energy with doping concentration 
provides an explanation for the observed decrease in unison of 
the spin-wave stiffness constant $\rho_s$ and the N\'{e}el 
temperature $T_{\rm N}$  with Ce doping in both
$\rm Nd_{2-x}Ce_x Cu O_4$ and $\rm Pr_{2-x}Ce_x Cu O_4$.\cite{electron3}
A spin dilution model, within which a Cu$^{2+}$ S=1/2 ion is assumed to 
be converted into a localized Cu$^{1+}$ S=0 ion for each added Ce ion, 
has been found to quantitatively describe the $x$-dependence of
$\rho_s$ and $T_{\rm N}$.\cite{electron2}
However, the localization of the added electrons in this scenario 
contradicts the metallic conductivity resulting from Ce doping.

\section{Spin-fluctuation correction to sublattice magnetization}
Figure 13 shows the first-order spin-fluctuation 
correction to the electron propagator.
Physically, the process 1 (2) represents a spin-$\uparrow$ electron
in state $\bf k$ below (above) the Fermi energy emitting a virtual magnon
with momentum $\bf Q$ and flipping into a spin-$\downarrow$ electron 
in state 

\begin{figure}
\vspace*{-130mm}
\hspace*{-43mm}
\psfig{file=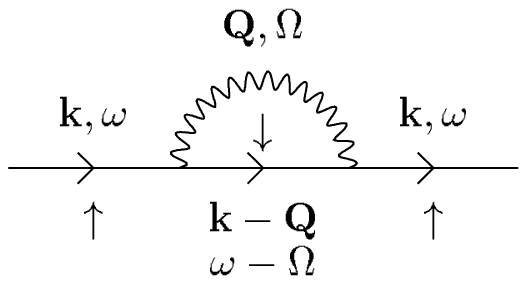,width=185mm,angle=0}
\vspace{-110mm}
\caption{
The quantum spin-fluctuation correction to the electron propagator,
representing a spin flip accompanied by the 
emission and absorption of a virtual magnon.}
\end{figure}

\noindent
${\bf k-Q}$ above (below) the Fermi energy, 
which then reabsorbs the virtual magnon. 
Such spin-flip processes involve transfer of spectral weight 
across the Fermi energy, and therefore result in corrections
to electron densities $n_\uparrow$ and $n_\downarrow$,
and hence to the sublattice magnetization.

Considering first the process (1) involving
$E_{\bf k} < E_{\rm F}$ and $E_{\bf k-Q} > E_{\rm F}$,
the integrated spectral weight transferred above the Fermi energy 
yields the reduction in the spin-$\uparrow$ density
on an A-sublattice site
\begin{eqnarray}
-\delta n_{\uparrow} ^{(1)} &=&
\sum_{\bf k} \int_{E_{\rm F}} ^\infty \frac{d\omega}{\pi}
{\rm Im} \; 
[\delta G_\uparrow ({\bf k}, \omega)]_{AA} \nonumber \\
&=&
\sum_{\bf k,Q} 
U^2 \int \frac{d\Omega}{\pi}
{\rm Im}\; \chi^{-+}_{\rm R}({\bf Q},\Omega)
\frac{a_{\bf k-Q,\downarrow} ^2}
{(E_{\bf k-Q} - E_{\bf k} + \Omega)^2} \nonumber \\
\end{eqnarray}
involving the retarded part of the magnon propagator
$\chi^{-+}_{\rm R}({\bf Q},\Omega)$.
Here $a_{\bf k-Q,\downarrow} ^2$ is 
the spin-$\downarrow$ electron density on the A sublattice site,
and in the strong coupling limit,
is given by 1 and $\epsilon_{\bf k-Q}^2 /4\Delta^2$ 
for state ${\bf k-Q}$ in the upper and lower band, respectively. 

For the undoped antiferromagnet,
where the spin-flip process involves an interband excitation, 
and states ${\bf k} $ and ${\bf k-Q} $ are energetically separated by 
the AF gap $2\Delta$, this correction reduces the A-sublattice density 
$n_\uparrow $ from 1 to 0.8 for the square lattice.\cite{spfluc}
A similar process transfers spin-$\downarrow$ spectral weight from
the upper (unoccupied) band to the lower band,
increasing  $n_\downarrow $ from 0 to 0.2,
resulting in a net reduction of 0.4 in the sublattice magnetization.

In the metallic antiferromagnet, however, spin-flip processes involve
the intraband excitations as well, and the small energy denominator
$E_{\bf k-Q}^\ominus - E_{\bf k}^\ominus \sim 4t'Q a$ in the $Q << 1$ limit
drastically increases the spin-fluctuation contribution to the
particle density correction.  On the other hand, 
the phase-space restriction on states ${\bf k} $ and ${\bf k-Q} $ 
to lie across the Fermi energy 
suppresses the intraband contribution in the long-wavelength limit
by a factor $Q$. In the following, 
we examine the combined effect of these two 
features of intraband excitations on the quantum correction to sublattice magnetization. 

As discussed earlier, except for a renormalized magnon velocity $c$
and amplitude $Z$, the coherent part of the 
magnon propagator for $Q<<1$ is qualitatively unchanged 
in the weak doping limit ($\sqrt{2}J > 4t'a$), 
with ${\rm Im}\; \chi^{-+}_{\rm R}({\bf Q},\omega) \sim (Z/Q) 
\delta (\omega - c Q)$ as for the AF insulator.
Substituting in Eq. (17), we obtain
\begin{eqnarray}
-\delta n_{\uparrow} ^{(1)} &=& 
\frac{U^2 t^2 a^4 }{4\Delta^2} 
\int \frac{Q \; dQ}{2\pi}
\int_{-\pi/2} ^{\pi/2}
\frac{a d\theta \; Q \cos \theta}{(2\pi)^2}
\cos ^2 2\theta  \nonumber \\
&\times & \int \frac{d\Omega}{\pi}
\frac{Z}{Q} \delta (\Omega - c Q)
\frac{1}
{(4t'a Q \cos \theta  + \Omega)^2} 
\end{eqnarray}
After doing the $\Omega$ integral,
the resulting $Q$ integral in Eq. (18) is of the form $\int dQ/Q$, 
which yields a  logarithmically divergent reduction in 
the A-sublattice density $n_\uparrow $ from process (1).
Similarly, the process (2) involving
$E_{\bf k}^\ominus > E_{\rm F}$ and $E_{\bf k-q}^\ominus < E_{\rm F}$, 
and the advanced part of the magnon propagator
$\chi^{-+}_{\rm A}({\bf Q},\Omega)$,
transfers spectral weight below the Fermi energy,
yielding an identical logarithmically divergent
enhancement to $n_\uparrow $.
Remarkably, it is this exact cancellation of two logarithmically divergent 
contributions which is responsible for the survival of long-range
AF order in the weakly (electron) doped cuprate. 

\section{Conclusions}
The study of spin fluctuations in the $t-t'$ Hubbard model shows that
when doped electrons (holes) are added to the bottom (top) of the upper (lower) 
Hubbard band, the consequent intraband particle-hole excitations   
strongly renormalize the magnon propagator in the metallic AF state.
The change in sign of the intraband coefficient $\alpha_{\rm intra}(\omega)$
with $\omega$ is a key result, 
leading to differences between static and dynamical behaviours.
Several features of the magnetic properties of the electron and hole doped cuprates
are understandable within this microscopic model. These include:
i)  finite (nearly zero) critical doping concentration $x_c$ above which 
long-range AF order is destroyed in electron (hole) doped cuprates, 
ii) a linear decrease in the spin-wave stiffness constant and 
N\'{e}el temperature with doping concentration, 
iii) magnon broadening and softening in hole doped cuprates,
iv) enhanced correlations at the dynamical level.

However, important features of the hole-doped cuprates require further investigation.
While the $t-t'$ Hubbard model (with negative $t'$) indicates instability of the 
AF state and tendency towards incommensurate ordering
for any finite hole doping, spin fluctuations in the hole-doped cuprates are
commensurate for low doping and a commensurate-incommensurate transition occurs at
$x \approx 0.05$. The exact mechanism responsible for the loss of long-range AF order
for $x < 0.05$ therefore remains unclear. 
Interestingly, it is in this doping regime that a magnon damping term $\Gamma \sim T$ 
in $\rm La_{2-x} Sr_x CuO_4$ and $\rm La_{2-x} Ba_x CuO_4$ 
has been observed at finite temperature,\cite{gamma,keimer}
which accounts for the anomalous nuclear spin relaxation rate, resistivity etc.

\end{document}